\documentclass[showpacs,preprintnumbers,amsmath,amssymb]{revtex4}
\usepackage{hyperref}
\usepackage{makeidx,endnotes,fancyhdr}
\usepackage{showidx}
\usepackage{amsmath,amssymb,theorem,epsf,eepic}

\newcommand{\ket}[1]{|#1\rangle}  
\def\kk{\rangle\!\rangle}\def\bb{\langle\!\langle}
\def\Tr{\operatorname{Tr}}

\def\>{\rangle}\def\<{\langle}
\def\sH{\mathcal{H}}

\def\map#1{{\mathcal #1}}\def\set#1{{\sf #1}}
\def\geq{\geqslant}

\begin{document}

\title{Optimal Time-Reversal of Multi-phase Equatorial States}  
\author{Francesco
  Buscemi}\email{buscemi@fisicavolta.unipv.it} \author{Giacomo Mauro
  D'Ariano}\email{dariano@unipv.it} \author{Chiara
  Macchiavello}\email{chiara@unipv.it} \affiliation{QUIT Group,
  Dipartimento di Fisica ``A.  Volta'', Universit\`a di Pavia, via
  Bassi 6, I-27100 Pavia, Italy}\homepage{http://www.qubit.it/}
\date{September 26, 2005}
\begin{abstract} Even though the time-reversal is unphysical (it
  corresponds to the complex conjugation of the density matrix), for
  some restricted set of states it can be achieved unitarily,
  typically when there is a common de-phasing in a $n$-level system.
  However, in the presence of multiple phases (i. e. a different
  de-phasing for each element of an orthogonal basis occurs) the time
  reversal is no longer physically possible. In this paper we derive
  the channel which optimally approaches in fidelity the time-reversal
  of multi-phase equatorial states in arbitrary (finite) dimension. We
  show that, in contrast to the customary case of the Universal-NOT on
  qubits (or the universal conjugation in arbitrary dimension), the
  optimal phase covariant time-reversal for equatorial states is a
  nonclassical channel, which cannot be achieved via a
  measurement/preparation procedure. Unitary realizations of the
  optimal time-reversal channel are given with minimal ancillary
  dimension, exploiting the simplex structure of the optimal maps.
\end{abstract}

\maketitle


\section{introduction}

Time reversal is not a physically achievable transformation on
arbitrary quantum states, since it corresponds to a positive, but not
completely positive map \cite{Kraus}. However, if restricted to
special sets of states, time reversal can be easily achieved by a
unitary transformation, e. g. when generating NMR spin-echoes by
turning over the fixed magnetic field along the rotation axis, with
all the spins rotating in the equatorial plane. For a general
$n$-level system, this is possible only when the levels are equally
spaced, corresponding to a common de-phasing. However, more generally,
when the levels are not equally spaced, we have different de-phasing
for each level, and in the presence of multiple phases the time
reversal is no longer physically possible.

The ideal time-reversal of a generic state corresponds to the
complex-conjugation (or, equivalently, transposition) of the
corresponding density matrix. Such transformation has also recently
attracted much interest in relation to the problem of entanglement, in
regards to the so-called PPT (positive partial transpose) criterion
\cite{peres,H3}. Since complex-conjugation cannot be achieved
unitarily, one can try to approximate the transformation with a
physical channel, optimizing the fidelity of the output state with the
complex-conjugated input. For the set of all pure states the resulting
optimal channel is "classical" \cite{univ-not,opt-transp}, in the
sense that it can be achieved by state-estimation followed by
state-preparation. In this paper we show that for multi-phase
equatorial states the optimal phase covariant time-reversal for
equatorial states is a nonclassical channel, namely it cannot be
achieved via the measurement/preparation procedure.  We will see that
the optimal channels form a simplex (i. e. a convex set which is
generated by convex combination of a finite set of extremal points, e.
g. a tetrahedron). Such a structure simplifies the search for unitary
realizations of the channels, which will be derived in the following
for minimal ancillary dimension.

The paper is organized as follows. In Section \ref{optimal-map} we
introduce the notation, and derive the optimal multi-phase conjugation
maps and the corresponding fidelity.  In Section \ref{unot-vs-phconj}
we compare the present optimal phase-covariant maps with the
universally covariant ones, and discuss their relation with optimal
state-estimation and phase-estimation.  In Section
\ref{convex-unitary} we analyze the simplex structure of the set of
optimal multi-phase conjugation maps, and explicitly construct their
unitary realizations with minimal ancilla dimension. Section
\ref{conclusions} closes the paper with some concluding remarks.


\section{optimal multi-phase conjugation maps}\label{optimal-map}
In the following we will restrict attention to equatorial states of a
$d$-dimensional quantum system, defined as
\begin{equation}\label{equat}
|\psi(\{\phi_j\})\>=\frac{1}{\sqrt d}(\ket{0}+e^{i\phi_1}\ket{1}
+e^{i\phi_2}\ket{2}+...+e^{i\phi_{d-1}}\ket{d-1}),
\end{equation}
expanded with respect to the fixed orthonormal basis
\begin{equation}\label{basis}
\set{B}\doteq\{|0\>,|1\>,\cdots,|d-1\>\}.
\end{equation}
of the Hilbert space $\sH$ of the quantum system. We consider
transformations that treat all input states (\ref{equat}) in the same
way, namely that are covariant under the group
$\mathbb{U}(1)^{\times(d-1)}$ of rotations of $d-1$ independent phases
$\{\phi_j\}$. The state (\ref{equat}) can be equivalently written as
\begin{equation}\label{seed}
|\psi(\{\phi_j\})\>=U(\{\phi_j\})|\psi_0\>,
\end{equation}
where 
\begin{equation}\label{eq:psi0}
|\psi_0\>=d^{-1/2}\sum_{i=0}^{d-1}|i\>
\end{equation}
is a fixed real state and
\begin{equation}
U(\{\phi_j\})=|0\>\<0|+\sum_{j=1}^{d-1}e^{i\phi_j}|j\>\<j|
\end{equation}
is the generic element of $\mathbb{U}(1)^{\times(d-1)}$.

We now derive the channel---e. g. completely positive trace-preserving
map---$\map{T}$ which optimally approximates the (anti-linear)
multi-phase conjugation on equatorial states, namely
\begin{equation}\label{exact-conj}
|\psi(\{\phi_j\})\>=\frac{1}{\sqrt{d}}(|0\>+e^{i\phi_1}|1\>+e^{i\phi_2}|2\>+\cdots)\longmapsto|\psi^*(\{\phi_j\})\>=\frac{1}{\sqrt{d}}(|0\>+e^{-i\phi_1}|1\>+e^{-i\phi_2}|2\>+\cdots).
\end{equation}
The map $\map{T}$ is covariant under the multi-phase
$\mathbb{U}(1)^{\times(d-1)}$ group, namely
\begin{equation}\label{covariance}
\map{T}\left(U(\{\phi_j\})\ \rho\ U^\dag(\{\phi_j\})\right)=U^*(\{\phi_j\})\ \map{T}(\rho)\ U^T(\{\phi_j\}).
\end{equation}
In the above equation $O^*$ ($O^T$) denotes the complex conjugation
(transposition) of the operator $O$ with respect to the orthonormal
basis (\ref{basis}) kept as real. Among all completely positive
trace-preserving (CPT) maps satisfying the covariance condition
(\ref{covariance}), we single out those maps which maximize the
fidelity between the output state and the ideally transformed state in
Eq. (\ref{exact-conj})
\begin{equation}\label{merit}
F\Big(|\psi^*(\{\phi_j\})\>,\map{T}(|\psi(\{\phi_j\})\>\<\psi(\{\phi_j\})|)\Big)=\Tr\Big[|\psi^*(\{\phi_j\})\>\<\psi^*(\{\phi_j\})|\ \map{T}(|\psi(\{\phi_j\})\>\<\psi(\{\phi_j\})|)\Big].
\end{equation}

Following Ref. \cite{non-univ}, we solve the optimization problem
under the covariance condition (\ref{covariance}), using the positive
operator $R$ on $\sH\otimes\sH$ defined as
\begin{equation}
R=(\map{T}\otimes\map{I})|\openone\kk\bb\openone|,
\end{equation}
where $\map{I}$ is the identity channel, and
$|\openone\kk=\sum_{i=0}^{d-1}|i\>|i\>$ is the maximally entangled
vector on $\sH\otimes\sH$ relative to the orthonormal basis $\set{B}$
in Eq. (\ref{basis}). The correspondence $\map{T}\leftrightarrow R$ is
one-to-one and can be inverted as
\begin{equation}\label{reconstruct}
\map{T}(\rho)=\Tr_{\sH_2}[\openone\otimes\rho^T\ R].
\end{equation}
In terms of the operator $R$ the trace-preservation condition for $\map{T}$ reads
\begin{equation}
\Tr_{\sH_1}[R]=\openone
\end{equation}
and the covariance property of the map (\ref{covariance}) becomes the
invariance for $R$
\begin{equation}\label{commutation}
[R,U^*(\{\phi_j\})^{\otimes 2}]=0.
\end{equation}
This, in turn, via the Schur Lemma implies the following block form
for $R$
\begin{equation}
R=\bigoplus_{\alpha:\textrm{ equiv. classes}}R_\alpha.
\end{equation}
The index $\alpha$ runs over the equivalence classes of irreducible
one-dimensional representations of $U(\{\phi_j\})^{\otimes 2}$
(without loss of generality we suppressed the complex conjugation).
The equivalence classes with the respective characters are listed in
Table \ref{table1}.
\begin{table}[h]
\begin{center}
\begin{tabular}{c|c}
Equivalence Classes & Characters\\
\hline \hline
$|00\>$ & 1\\
$|11\>$ & $e^{2i\phi_1}$\\
$\vdots$ & $\vdots$\\
$|ii\>$ & $e^{2i\phi_i}$\\
$\vdots$ & $\vdots$\\
$|01\>,|10\>$ & $e^{i\phi_1}$\\
$\vdots$ & $\vdots$\\
$|ij\>,|ji\>,\quad i\neq j$ & $e^{i(\phi_i+\phi_j)},\quad i\neq j$\\
$\vdots$ & $\vdots$\\
\end{tabular}
\end{center}
\caption{Equivalence classes and respective characters of irreducible
one-dimensional representations of $U(\{\phi_j\})^{\otimes 2}$.\label{table1}}
\end{table}

Noticing that $|\psi_0\>\<\psi_0|=|\psi_0\>\<\psi_0|^*$, and using
Eqs.~(\ref{seed}),~(\ref{reconstruct}) and~(\ref{commutation}), we can
rewrite the fidelity in Eq.~(\ref{merit}) as
\begin{equation}\label{fidelity}
F=\Tr[|\psi_0\>\<\psi_0|^{\otimes 2}\ R].
\end{equation}
From Eq.~(\ref{eq:psi0}), one has $|\psi_0\>\<\psi_0|^{\otimes 2}=
d^{-2}\sum_{i,j,k,l}|i\>\<j|\otimes|k\>\<l|$, and the
fidelity~(\ref{fidelity}) is equal to
$F=d^{-2}\sum_\alpha\sum_{i,j,k,l} r^{(\alpha)}_{jl,ik}$, where
$r^{(\alpha)}_{jl,ik}=\<jl|R_\alpha|ik\>$. As argued in
Refs.~\cite{DaMa,BuDaMa}, the maximum $F$ is then obtained when the
off-diagonal terms of the operator $R$ are positive and as large as
possible, namely when the blocks $R_\alpha$ are
rank-one~\cite{note-on-positivity}, or in equations
\begin{equation}
R=\frac{1}{\mathcal{N}}\left[\sum_ic_i|ii\>\<ii|+\sum_{i>j}c_{ij}(\alpha_{ij}|ij\>+\beta_{ij}|ji\>)(\alpha^*_{ij}\<ij|+\beta^*_{ij}\<ji|) \right],
\end{equation}
where $\mathcal{N}=(\sum_ic_i+\sum_{i>j}c_{ij})/d$ is a normalization
constant, $|\alpha_{ij}|^2+|\beta_{ij}|^2=1$, and $c_i\geq0$,
$c_{ij}\geq0$, $\forall$ $i,j$.  The fidelity then takes the form
\begin{equation}
\begin{split}
F&=\Tr[|\psi_0\>\<\psi_0|^{\otimes 2}\ R]\\
&=\frac{1}{d^2}\ \frac{d}{\sum_ic_i+\sum_{i>j}c_{ij}}\ \left[\sum_ic_i+\sum_{i>j}c_{ij}\left(|\alpha_{ij}|^2+|\beta_{ij}|^2+2\Re(\alpha_{ij}\beta_{ij}^*)\right)\right]\\
&=\frac{1}{d}\ \frac{1}{\sum_ic_i+\sum_{i>j}c_{ij}}\ \left[\sum_ic_i+\sum_{i>j}c_{ij}\left(1+2\Re(\alpha_{ij}\beta_{ij}^*)\right)\right]\\
&=\frac{1}{d}+2\frac{\sum_{i>j}c_{ij}\Re(\alpha_{ij}\beta_{ij}^*)}{d\left(\sum_ic_i+\sum_{i>j}c_{ij}\right)}.
\end{split}
\end{equation}
Now, the maximum of $\Re(\alpha_{ij}\beta_{ij}^*)$ is achieved when
$\alpha_{ij}=\beta_{ij}$, implying $|\alpha_{ij}|^2=1/2$ for all
$i,j$. Therefore we choose $\alpha_{ij}=2^{-1/2}$. The fidelity
becomes
\begin{equation}
F=\frac{1}{d}+\frac{\sum_{i>j}c_{ij}}{d\left(\sum_ic_i+\sum_{i>j}c_{ij}\right)}.
\end{equation}
In order to maximize $F$ we put $c_i=0$ for all $i$. The optimal
fidelity takes the following simple form
\begin{equation}\label{opt-fid}
F=\frac{2}{d}.
\end{equation}
Then, we impose the trace-preservation condition, obtaining
\begin{equation}\label{trace-preserv}
\begin{split}
\Tr_1[R]&=\frac{d}{2\sum_{i>j}c_{ij}}\sum_{i>j}c_{ij}(|i\>\<i|+|j\>\<j|)\\
&\doteq\sum_{i>j}b_{ij}(|i\>\<i|+|j\>\<j|)\\
&\equiv\openone,
\end{split}
\end{equation}
where 
\begin{equation}
b_{ij}=d\frac{c_{ij}}{2\sum_{i>j}c_{ij}}.
\end{equation}
Since the projector $|0\>\<0|$, for example, appears in the sum
(\ref{trace-preserv}) multiplied by $\sum_{i=1}^{d-1}b_{i0}$, the
coefficients are constrained as follows
\begin{equation}
\sum_{i=1}^{d-1}b_{i0}=1,
\end{equation}
and, similarly, for the $|j\>\<j|$, element one has
\begin{equation}\label{bistoch}
\sum_{i=0}^{j-1}b_{ji}+\sum_{i=j+1}^{d-1}b_{ij}=1.
\end{equation}
Rearranging the positive coefficients $\{b_{ij}\}_{i>j}$ into a square
matrix array, they define the lower-trianguar section of a square
matrix.  Such a matrix can be \emph{uniquely} completed to a null-diagonal
symmetric bistochastic matrix, that is, a symmetric null-diagonal
matrix with non-negative entries, such that all its rows' and columns'
entries sum up to one, namely all its rows and columns are probability
distributions.

Up to now the operator $R$ is simplified as follows
\begin{equation}\label{optimal-R}
R=\sum_{i>j}b_{ij}(|ij\>+|ji\>)(\<ij|+\<ji|),
\end{equation}
where the coefficients $b_{ij}$'s, uniquely defining a map $\mathcal
T$ achieving the optimal fidelity $2/d$, are the entries of a
null-diagonal symmetric bistochastic matrix. From
Eq.~(\ref{optimal-R}) and the reconstruction
formula~(\ref{reconstruct}), an optimal phase covariant transposition
map can then be easily expressed in the Kraus form as follows
\begin{equation}
\mathcal{T}(\rho)=\sum_{i>j}b_{ij}(|i\>\<j|+|j\>\<i|)\ \rho\ (|i\>\<j|+|j\>\<i|).
\end{equation}

Notice that the constraint (\ref{bistoch}) over $\{b_{ij}\}$ is indeed
very strong: for qubits and qutrits it suffices to completely and
univocally determine the map.  In the case of qubits the only
null-diagonal symmetric bistochastic matrix is
\begin{equation}\label{qubit}
\{b_{ij}\}=\begin{pmatrix}
  0 & 1\\
  1 & 0
\end{pmatrix},
\end{equation}
and the optimal transposition map is the unitary transformation
\begin{equation}\label{unitary-qubits}
\mathcal{T}_2(\rho)=\sigma_x\rho\sigma_x\equiv\rho^*,
\end{equation}
which clearly achieves F=1. For $d=3$ there is again a unique choice for a null-diagonal symmetric
bistochastic matrix, which is given by
\begin{equation}\label{qutrit}
\{b_{ij}\}=\begin{pmatrix}
0 & 1/2 & 1/2\\
1/2 & 0 & 1/2\\
1/2 & 1/2 & 0
\end{pmatrix}.
\end{equation}
For $d\geq4$ there exist many optimal maps. For example, for
$d=4$ we can parametrise the family of maps by varying two positive
parameters $p_1$ and $p_2$ such that $0\le p_1+p_2=p_{12}\le 1$:
\begin{equation}\label{NSB-4}
\{b_{ij}\}=\begin{pmatrix}
0 & p_1 & p_2 & 1-p_{12}\\
p_1 & 0 & 1-p_{12} & p_2\\
p_2 & 1-p_{12} & 0 & p_1\\
1-p_{12} & p_2 & p_1 & 0
\end{pmatrix}.
\end{equation}


\section{universal transposition and multi-phase conjugation}
\label{unot-vs-phconj}

In Refs. \cite{univ-not,opt-transp} it was shown that the fidelity of
the optimal universal transposition map is equal to the fidelity of
the optimal universal pure state estimation \cite{MaPo,BruMa}, that
for a single input copy is given by
\begin{equation}\label{univ}
F=\frac{2}{d+1},
\end{equation}
which is always lower than the fidelity of optimal phase covariant
transposition (\ref{opt-fid}), as expected. The equivalence between
transposition and state estimation means that an optimal universal
transposition can be achieved by optimally estimating the input state
and then preparing the transposed state. In this sense the optimal
universal transposition is a "classical" map.

In contrast to the universal case, the phase covariant transposition map cannot be achieved by phase
estimation/preparation. Indeed, the fidelity of optimal multi-phase estimation for a single input
copy is given by \cite{Ma} 

\begin{equation}
F=\frac{2d-1}{d^2},
\end{equation}
which is always smaller that the optimal fidelity of the phase
covariant transposition map (\ref{opt-fid}). Hence, the optimal phase
covariant transposition is a genuinely quantum channel. The situation
is particularly striking in the case of qubits, where it is possible
to perfectly transpose all equatorial states, while the phase can
never be measured exactly with finite resources \cite{DeBuEk}.


\section{convex structure and physical realization of optimal maps}
\label{convex-unitary}

In Section \ref{optimal-map} we have shown that optimal multi-phase
conjugation maps are in one-to-one correspondence with null-diagonal
symmetric bistochastic (NSB) matrices, which form a convex set. On the
other hand, every bistochastic matrix is a convex combination of
permutation matrices---this is the content of the Birkhoff theorem
\cite{Bh}. The null-diagonal and symmetry constraints, however, force
the convex set of NSB matrices to be strictly contained into the
convex polyhedron of bistochastic matrices. This fact causes the
extremal NSB matrices to eventually lie strictly inside the set of
bistochastic matrices, generally preventing them from being
permutations.

The geometrical study of the set of NSB matrices and its extremal
points can shed some light on the unusual feature that there exist
different "equally optimal" maps. The problem arises for dimension
at least $d=4$. In this case the decomposition of the matrix
$\{b_{ij}\}$ in Eq. (\ref{NSB-4}) into extremal components is
\begin{equation}\label{ext-dec-4}
\{b_{ij}\}=p_1\begin{pmatrix}
0& 1& 0& 0\\
1& 0& 0& 0\\
0& 0& 0& 1\\
0& 0& 1& 0
\end{pmatrix}+p_2\begin{pmatrix}
0& 0& 1& 0\\
0& 0& 0& 1\\
1& 0& 0& 0\\
0& 1& 0& 0
\end{pmatrix}+p_3\begin{pmatrix}
0& 0& 0& 1\\
0& 0& 1& 0\\
0& 1& 0& 0\\
1& 0& 0& 0
\end{pmatrix}=p_1P^{(1)}+p_2P^{(2)}+p_3P^{(3)},
\end{equation}
where $p_1,p_2,p_3\geq 0$ and $p_1+p_2+p_3=1$. A natural question is
now which optimal maps can be achieved with minimal resources.

In order to physically realize a given CPT map $\mathcal{E}$, one
needs to design a specific unitary interaction $U$ and prepare an
ancilla in a specific state, say $|0\>\<0|_a$, in such a way that
\begin{equation}
\mathcal{E}(\rho)=\Tr_a[U\;(\rho\otimes|0\>\<0|_a)\;U^\dag].
\end{equation}
This is always possible \cite{Kraus,BuDaSa}. The existence of
equivalently optimal maps allows us to choose between realizations
with either a smaller ancilla dimension, or a more flexible ancilla
state-preparation.

More explicitly, for $d=4$, we define three unitaries $U_1$, $U_2$ and
$U_3$ on $\mathbb{C}^4\otimes\mathbb{C}^2$ as
\begin{equation}\label{unitaries-for-4}
U_1\doteq\begin{pmatrix}
T_{10} & T_{32}\vspace{0.2cm}\\
T_{32} & T_{10}
\end{pmatrix},\quad
U_2\doteq\begin{pmatrix}
T_{20} & T_{31}\vspace{0.2cm}\\
T_{31} & T_{20}
\end{pmatrix},\quad
U_3\doteq\begin{pmatrix}
T_{30} & T_{21}\vspace{0.2cm}\\
T_{21} & T_{30}
\end{pmatrix},
\end{equation}
where $T_{ij}=|i\>\<j|+|j\>\<i|$. Each of them realizes an extremal
optimal multi-phase conjugation map (corresponding to $p_k=1$ in Eq.
(\ref{ext-dec-4}) for a given $k$), namely
\begin{equation}\label{ext-unit-4}
  \mathcal{T}^{(k)}_4(\rho)=\sum_{i>j}P^{(k)}_{ij}T_{ij}\rho T_{ij}=\Tr_a[U_k\;(\rho\otimes|0\>\<0|_a)\;U_k^\dag],
\end{equation}
where $|0\>\<0|_a$ is a fixed qubit ancilla state. Notice that the
ancilla must not necessarily be in a pure state, and the optimal map
is equivalently achieved for diagonal mixed ancilla state
$\alpha|0\>\<0|_a+\beta|1\>\<1|_a$. By adding a control qutrit, we can
now choose among any of the optimal maps using the controlled-unitary
operator on $\mathbb{C}^4\otimes\mathbb{C}^2 \otimes\mathbb{C}^3$
\begin{equation}
U=U_1\otimes|0\>\<0| + U_2\otimes|1\>\<1|
+U_3\otimes|2\>\<2|.
\end{equation}
Any optimal multi-phase conjugation map can now be written as
\begin{equation}\label{unitary-4}
\mathcal{T}_4(\rho)=\Tr_{a,b}\left[U\;\left(\rho\otimes|0\>\<0|_a\otimes\sigma_b\right)\;U^\dag\right]
\end{equation}
where $\sigma_b$ is a generic density matrix on $\mathbb{C}^3$. By
superimposing or mixing the three orthogonal states
$\{|0\>,|1\>,|2\>\}$ of the qutrit we control the weights
$p_1,p_2,p_3$ in Eq. (\ref{ext-dec-4}) via the diagonal entries of
$\sigma_b$.

Eqs. (\ref{unitaries-for-4})-(\ref{unitary-4}) can be generalized for
higher even dimensions \cite{odd-dim}, with
\begin{equation}
\begin{split}
&U_k\doteq\sum_{i,j=0}^{\frac{d}{2}-1}T_{k\oplus 2i\oplus 2j,2i\oplus 2j}\otimes|i\>\<j|,\qquad k=1,\dots,d-1,\\
&U=\sum_{k=1}^{d-1}U_k\otimes|k\>\<k|,\\
&\mathcal{T}^{(k)}_d(\rho)=\Tr_a[U_k\;(\rho\otimes|0\>\<0|_a)\;U_k^\dag],\\
&\mathcal{T}_d(\rho)=\Tr_{a,b}\left[U\;\left(\rho\otimes|0\>\<0|_a\otimes\sigma_b\right)\;U^\dag\right]\\
\end{split}
\end{equation}
where $U_k$'s are unitary operators acting on
$\mathbb{C}^d\otimes\mathbb{C}^{d/2}$, $U$ is a control-unitary
operator on
$\mathbb{C}^d\otimes\mathbb{C}^{d/2}\otimes\mathbb{C}^{d-1}$,
$|0\>\<0|_a$ is a fixed $(d/2)$-dimensional pure state, and $\sigma_b$
is a generic $(d-1)$-dimensional density matrix. The minimum dimension
of the ancilla space required to unitarily realize an optimal phase
covariant transposition map is $d/2$, generalizing the result for
$d=4$, for which just a qubit is needed (see Eq. (\ref{ext-unit-4})).
Notice that realization of phase covariant transposition generally
needs much less resources than realization of universal transposition:
the dimension $d/2$ of the ancilla space in the phase covariant case
has to be compared with the dimension $d^2$ required in the universal
case \cite{opt-transp}.

As a final remark, notice that unitary realizations of CPT channels
are far from being uniquely determined \cite{BuDaSa}: here we chose
the controlled unitary structure because of its clear geometrical
interpretation in connection with the convex structure of the
polyhedron of optimal maps.


\section{conclusions}\label{conclusions}

In this paper we have derived the channel which optimally approaches
in fidelity the time-reversal of multi-phase equatorial states in
arbitrary (finite) dimension. We show that, in contrast to the
customary case of the Universal-NOT on qubits (or the universal
conjugation in arbitrary dimension), the optimal phase covariant
time-reversal for equatorial states is a nonclassical channel, which
cannot be achieved via a measurement/preparation procedure. We have
given unitary realizations of the optimal time-reversal channel with
minimal ancillary dimension, exploiting the simplex structure of the
optimal maps. The optimal channels are related to null-diagonal
symmetric bistochastic matrices. For $d\geq4$ this gives a simplex
structure of equivalently optimal maps.


\appendix
\section*{Acknowledgements}

This work has been jointly funded by the EC under the programs ATESIT
(Contract No. IST-2000-29681), SECOQC (Contract No. IST-2003-506813)
and INFM PRA-CLON.




\begin{thebibliography}{}
\bibitem{odd-dim} The case of odd dimensions is much more complicated
  and will be not analysed here. The problem with odd dimensions is
  that extremal points of the convex set of NSB matrices are not
  permutations. Hence Birkhoff theorem cannot be applied.
\end{thebibliography}
\end{document}